\documentclass[showpacs,twocolumn,prl,aps]{revtex4}
\usepackage{amssymb}
\usepackage{amsmath}
\usepackage{graphicx}

\begin{document}

\title{Spin-orbit scattering in quantum diffusion of massive Dirac fermions}
\author{Wen-Yu Shan}
\affiliation{Department of Physics and Centre of Theoretical and Computational Physics,
The University of Hong Kong, Pokfulam Road, Hong Kong, China}
\author{Hai-Zhou Lu$^*$}
\affiliation{Department of Physics and Centre of Theoretical and Computational Physics,
The University of Hong Kong, Pokfulam Road, Hong Kong, China}
\author{Shun-Qing Shen$^{\dag}$}
\affiliation{Department of Physics and Centre of Theoretical and Computational Physics,
The University of Hong Kong, Pokfulam Road, Hong Kong, China}
\date{\today }

\begin{abstract}
Effect of spin-orbit scattering on quantum diffusive transport of two-dimensional massive Dirac fermions is studied by the diagrammatic technique.
The quantum diffusion of massive Dirac fermions can be viewed as a singlet Cooperon in the massless limit and a triplet Cooperon in the large-mass limit.
The spin-orbit scattering behaves like random magnetic fields only to the triplet Cooperon, and suppresses the weak localization of Dirac fermions in the large-mass regime.
This behavior suggests an experiment to detect the weak localization of bulk subbands in topological insulator thin films, in which a narrowing of the cusp of the negative magnetoconductivity is expected after doping heavy-element impurities. Finally, a detailed comparison between the conventional two-dimensional electrons and Dirac fermions is presented for impurities of orthogonal, symplectic, and unitary symmetries.
\end{abstract}

\maketitle

Recently, a featured observation in the quantum diffusive transport of topological insulators (TIs) \cite{Hasan10rmp,Qi11rmp,Moore10nat} is the weak antilocalization (WAL) \cite{Checkelsky09prl,Peng10NatMat,Chen10prl,Checkelsky11prl,He11prl,Liu11prb,Wang11prb,Liu12prl,Chen11rc,Kim11prb,Steinberg11prb},
an enhancement of conductivity due to destructive interference between time-reversed scattering loops.
It exhibits as a negative cusp in low-field magnetoconductivity or logarithmically decreasing conductivity with increasing temperature in the metallic regime.

The destructive interference is due to the $\pi$ Berry phase of massless Dirac fermions \cite{Ando98jpsj,Hsieh09nat}.
If Dirac fermions acquire a mass, the $\pi$ Berry phase will be changed, leading to the crossover from WAL to its opposite, the weak localization (WL) \cite{Imura09prb,Ghaemi10prl,Lu11prl,Dugaev01prb}.
Because both their surface and bulk states are Dirac fermions \cite{Zhang09np},
topological insulators provide a platform where rich quantum transport phenomena of Dirac fermions can be explored.
The WL-WAL crossover has been observed \cite{Liu12prl} for the magnetically doped surface states \cite{Hor10prb,Chen10sci,Wray11np}.
Suppressed WAL was also observed in thin films \cite{Liu11prb,Kim11prb}, probably due to the finite-size effect of the surface states \cite{Ghaemi10prl,Zhang10np}.
More interestingly, the bulk states of topological insulators always have the mass, which is nothing but the energy gap.
Unlike conventional electrons, where spin-orbit coupling always leads to WAL \cite{Bergmann84PhysRep,HLN80,Maekawa81jpsj,Bergman81prl},
the bulk states of topological insulator could give rise to WL due to their massive Dirac fermion nature \cite{Lu11prb}.

In the language of ``Cooperon" (the vertex of maximally crossed diagrams that gives the quantum interference correction to conductivity), the conventional two-dimensional electrons have two spin-1/2 bands crossing the Fermi surface, giving four ``Cooperon" channels: one ``singlet" and three ``triplets", depending on the total spin angular momentum of incoming and outgoing electrons. Each channel contributes a conductivity correction of the same magnitude but the singlet (triplet) gives WAL (WL). For only elastic scattering, the triplets outnumber the singlet by 2, leading to WL in total. On the other hand, strong spin-orbit scattering can suppress only the triplets due to spin relaxation, resulting in the crossover to WAL \cite{Bergmann84PhysRep,Millis84prb,Efetov80sov,Altshuler81sov}.
Therefore, the spin-orbit scattering may also suggest an approach to probe the WL of the bulk states of TIs. However,
it is still unclear what will happen to massive Dirac fermions if spin-orbit scattering is introduced.

\begin{table}[htbp]
\caption{``Cooperon" channels for the conventional electrons and Dirac fermions.
Triplet (singlet) channel gives WL (WAL).
Spin-orbit scattering only quenches the triplet channels, leading to the crossover from WL to WAL for conventional electrons and the suppression of WL in the large-mass limit of Dirac fermions. $\gamma$ is the Berry phase.  }
\label{tab:cooperoons}%
\begin{ruledtabular}
\begin{tabular}{ccc}
``Cooperon" channels&   ``triplet" &  ``singlet" \\
&  ($\Rightarrow$WL) &  ($\Rightarrow$WAL) \\
\hline
conventional electron&  $\times 3$ &  $\times 1$\\
massless Dirac fermion  ($\gamma=\pi$)  & &  $\times 1$    \\
large-mass Dirac fermion ($\gamma\rightarrow 0$) &   $\times 1$  \\
\end{tabular}
\end{ruledtabular}
\end{table}

In this work, we study the effects of spin-orbit scattering on the quantum diffusive transport of massive Dirac fermions.
For a single cone of Dirac fermions, the transport can be understood as a ``singlet Cooperon" in the massless limit and a ``triplet Cooperon" in the large-mass limit. The spin-orbit scattering has no effect on the ``singlet",
but behaves like random magnetic impurities to the ``triplet".
As a result, the spin-orbit scattering can drastically suppress the WL in the large-mass regime,
while has little impact on the WAL in the small-mass regime.
We expect a narrowing of the cusp of the negative magnetoconductivity of WAL, by doping impurities of heavy non-magnetic elements (e.g., Au) onto thin films of topological insulators.
Finally, a systematic comparison is presented for the conventional and Dirac fermions in the presence of scattering by ordinary, spin-orbit, and magnetic impurities.

Both the two-dimensional bulk and surface states of topological insulator can be described by the Hamiltonian of massive Dirac fermions,
\begin{eqnarray}\label{diracmodel}
H=\vec{d} \cdot \vec{\sigma} ,
\end{eqnarray}
where $\vec{d}=(\hbar v k_y, -\hbar vk_x, d_z)$, $\vec{\sigma}$ is the vector of Pauli matrices, $\hbar $ is Planck's constant over $2\pi$, and $v$ is the effective velocity. $d_z=\Delta/2-Bk^2$ gives the mass for the Dirac cones in the effective model of the two-dimensional bulk subbands \cite{Lu11prb} or the surface bands \cite{Lu10prb}, and $d_z=\Delta/2$ for magnetically doped surface states \cite{Hor10prb,Chen10sci,Wray11np}.
The spin-orbit scattering is described by
\begin{eqnarray}
U_{\mathrm{so}}(\bold{r})&=&\sum_i\frac{\hbar}{4m^2c^2}
\vec{\sigma}\cdot\nabla u(\bold{r}-\bold{R}_{i}^{\mathrm{so}})\times\bold{p},
\end{eqnarray}
where $u(\bold{r}-\bold{R}_{i}^{\mathrm{so}})$ represents the random potential by an impurity located at $\bold{R}_{i}^{\mathrm{so}}$.
The strength of the spin-orbit scattering will be characterized by a length $\ell_{\mathrm{so}}$.
Shorter $\ell_{\mathrm{so}}$ means stronger spin-orbit scattering.
$\ell_{\mathrm{so}}=\sqrt{D\tau_{\mathrm{so}}}$, where $D$ is the diffusion constant, $\tau_{\mathrm{so}}$ is the spin-orbit scattering time with $1/\tau_{\mathrm{so}}=2/\tau_{\mathrm{so},x}+1/\tau_{\mathrm{so},z}$,
\begin{eqnarray}\label{relaxation_time1}
\frac{1}{\tau_{\mathrm{so},z}}&=&
\frac{\pi N_F}{\hbar}(1+\cos^2\theta)n_{\mathrm{so}}u_{\mathrm{so},z}^2\overline{(\bold{k\times k^{'}})_z^2},\nonumber\\
\frac{1}{\tau_{\mathrm{so},x}}&=&\frac{\pi N_F}{\hbar}(1-\cos^2\theta )n_{\mathrm{so}}u_{\mathrm{so},x}^2\overline{(\bold{k\times k^{'}})_x^2},
\end{eqnarray}
where $N_F$ is the density of states at the Fermi energy, $n_{\mathrm{so}}$ is the concentration of impurities that induce spin-orbit scattering. $u_{\mathrm{so},x/z}$ are defined from $u(\bold{r}-\bold{R}_{i}^{\mathrm{so}})$ by assuming delta potentials, $\cos\theta\equiv d_F/\sqrt{d_F^2+(\hbar v k_F)^2}$, $d_F$ is the value of $d_z$ at the Fermi surface, $k_F$ is the Fermi wave vector. In above definitions, we have replaced $(\bold{k\times k^{'}})_i^2$ by its average over momentum directions $\overline{(\bold{k\times k^{'}})_i^2}$. This replacement is valid as long as $\ell_{\mathrm{so}}\gg \ell_e$ \cite{Rammer-book}, and $\ell_e$ is the length that characterizes the elastic scattering, which is always the strongest scattering mechanism in the quantum diffusive transport. Although intrinsic spin-orbit coupling carried by band dispersions of Dirac fermions and extrinsic spin-orbit scattering share the same origin at the atomic level, their effects on the spin relaxation \cite{Dresselhaus92prl,Iordanskii94jetplett} are quite different, similar to the difference between the Elliott-Yafet \cite{Elliott54pr,Yafet52pr} and Dyakonov-Perel mechanisms \cite{DP71jetp}.

With the help of the diagrammatic technique \cite{HLN80,Altshuler80prb,Bergmann84PhysRep,Suzuura02prl,Shon98jpsj,McCann06prl},
we calculated the quantum interference correction to conductivity and magnetoconductivity for the massive Dirac model
in the presence of elastic, spin-orbit, and magnetic scatterings.
The formulas in the presence of spin-orbit scattering remain the same two-term structure as the one with only elastic and magnetic scatterings \cite{Lu11prl}. The difference is the replacements $1/\tau_x\rightarrow 1/\tau_x -1/\tau_{\mathrm{so},x}$ and $1/\tau_z\rightarrow 1/\tau_z -1/\tau_{\mathrm{so},z}$, where $\tau_x$ and $\tau_z$ are the in-plane and out-of-plane magnetic scattering times, respectively. Besides, the total relaxation time now is given by $1/\tau=1/\tau_e+1/\tau_m+1/\tau_{\mathrm{so}}$, where the magnetic scattering time $1/\tau_m=2/\tau_x+1/\tau_z$ and the spin-orbit scattering time $1/\tau_{\mathrm{so}}=2/\tau_{so,x}+1/\tau_{so,z}$.

The new findings can be presented by reviewing the previous results.
The massive Dirac model in Eq. (\ref{diracmodel}) carries a Berry phase given by $\gamma=\pi(1-\cos\theta)$, which can be tuned from $\pi$ to 0 as $\cos\theta$ changes from 0 to 1, corresponding to the massless and large-mass limits, respectively. If there were only the elastic scattering, a single massless ($\gamma=\pi$) Dirac cone will exhibit WAL in the quantum diffusive transport. By introducing the mass term ($\gamma\neq \pi$), which changes the interference scenario from destructive to constructive due to the change of $\pi$ Berry phase to 0, a crossover from WAL to WL will be expected \cite{Lu11prl,Ghaemi10prl,Liu12prl}.

\begin{figure}[htbp]
\centering
\includegraphics[width=0.49\textwidth]{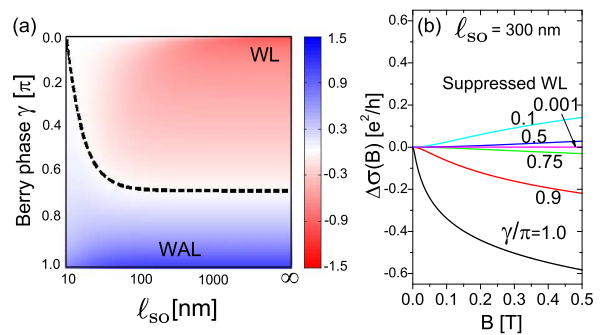}
\caption{(a) Zero-field quantum interference correction to conductivity $\sigma^F(0)$ (in units of $e^2/h$) as a function of $\gamma$ and the spin-orbit scattering length $l_{\mathrm{so}}$. The dashed curve separates the positive WAL and negative WL regimes. (b) Magnetoconductivity $\Delta\sigma(B)$ for different $\gamma$ at $\ell_{\mathrm{so}}=300$nm. Other parameters: elastic scattering length $\ell_e$=10nm, phase coherence length $\ell_{\phi}=300$ nm, and magnetic scattering length $\ell_{m}\rightarrow\infty$. }
\label{fig:diagram}
\end{figure}

We find that the spin-orbit scattering will suppress the weak localization in the large mass regime.
Fig. \ref{fig:diagram}(b) presents the magnetoconductivity.
WAL (WL) can be recognized as a sharp negative (positive) low-field magnetoconductivity cusp. By introducing a small spin-orbit scattering ($\ell_{\mathrm{so}}=300$ nm),
the negative curves of WAL with $\gamma/\pi$ =1, 0.9, 0.75 are almost unaffected while the positive curves with $\gamma$ =0.5, 0.1, 0.001 are suppressed drastically.
Even, the magnetoconductivity of WL totally vanishes in the $\gamma\rightarrow 0$ limit.
In Fig. \ref{fig:diagram}(a), we present the quantum interference conductivity correction $\sigma^F(0)$ as a function of $\gamma$ and $\ell_{\mathrm{so}}$.
WAL (WL) demonstrates as positive (negative) $\sigma^F(0)$ at low temperatures and zero field.
In the absence of spin-orbit scattering ($\ell_{\mathrm{so}} \rightarrow \infty$), $\sigma^F(0)$ reaches positive maximum (WAL) when $\gamma=\pi$ and negative minimum (WL) when $\gamma=0$. The boundary (dashed curve) that separates the positive WAL and negative WL regimes approaches to a saturate value of $\gamma\approx0.3$ \cite{Imura09prb} as $\ell_{\mathrm{so}}\rightarrow \infty$.
As we increase the spin-orbit scattering (decreasing $\ell_{\mathrm{so}}$), the WL regime shrinks. In the limit of strong spin-orbit scattering ($\ell_{\mathrm{so}}\rightarrow 0$), the regime of WL vanishes, and the WAL-WL crossover on the $\ell_{\mathrm{so}}\rightarrow \infty$ side is replaced by WAL .

\begin{figure}[htbp]
\centering
\includegraphics[width=0.49\textwidth]{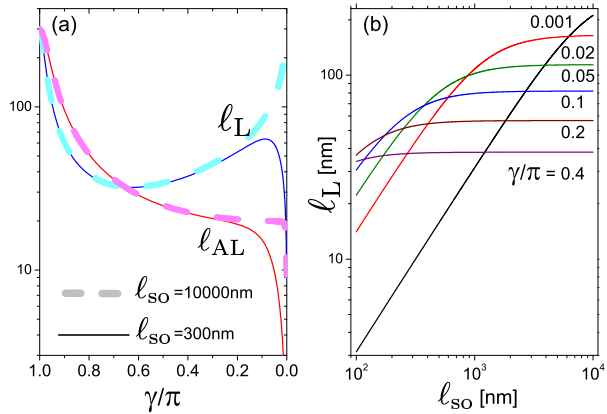}
\caption{(a) The weak antilocalization phase coherence length ($\ell_{\mathrm{AL}}$) and the weak localization phase coherence length ($\ell_{\mathrm{L}}$) as functions of $\gamma$ for weak ($\ell_{\mathrm{so}}=10000$nm) and strong ($\ell_{\mathrm{so}}=300$nm) spin-orbit scattering. (b) The weak localization phase coherence length $\ell_{\mathrm{L}}$ as a function of the spin-orbit scattering length $\ell_{\mathrm{so}}$ for different $\gamma$. }
\label{fig:LL}
\end{figure}

The suppression of WL by spin-orbit scattering can be further studied by examining the effective phase coherence lengthes.
In the quantum diffusion transport, an electron can be scattered by static centers for many times but still maintains its phase.
The phase, is protected by a long phase coherence length $\ell_{\phi}$ (set as 300 nm in this work according to experimental fittings).
In our calculated quantum interference correction to conductivity, the WL-WAL crossover can always be described by two competing terms, one for WL and the other for WAL. Each term looks like a Hikami-Larkin-Nagaoka formula \cite{HLN80}, and is characterized by an effective phase coherence length, denoted as $\ell_{\mathrm{L}}$ and $\ell_{\mathrm{AL}}$, respectively. For very long $\ell_{\mathrm{so}}$ (weak spin-orbit scattering), $\ell_{\mathrm{AL}}$ diverges as $\gamma\rightarrow \pi$, and $\ell_{\mathrm{L}}$ diverges as $\gamma\rightarrow 0$ [see dashed curves in Fig. \ref{fig:LL}(a)]. The divergence of effective phase coherence lengthes protects the WAL as $\gamma\rightarrow \pi$ and WL as $\gamma\rightarrow0$.
For short $\ell_{\mathrm{so}}$ (strong spin-orbit scattering), the WAL phase coherence length $\ell_{\mathrm{AL}}$ remains almost unchanged in the small mass regime. However, the divergence of $\ell_{\mathrm{L}}$ in the large mass regime is suppressed [the solid curves in Fig. \ref{fig:LL}(a)].
Fig. \ref{fig:LL}(b) shows $\ell_{\mathrm{L}}$ as a function of $\ell_{\mathrm{so}}$ in the large mass limit.
The slope of the curves shows that the spin-orbit scattering has stronger influence on larger mass cases.
In the $\gamma\rightarrow 0$ limit, $\ell_{\mathrm{L}}\rightarrow 0$ as $\ell_{\mathrm{so}}\rightarrow 0$, leading to the vanishing magnetoconductivity in Fig. \ref{fig:diagram}(b).

The physical picture of the suppression of WL can be understood by the ``Cooperons" \cite{Efetov80sov,Altshuler81sov}.
The quantum diffusion transport of a single cone of Dirac fermions can be understood as a ``singlet Cooperon" in the massless limit and a ``triplet Cooperon" in the large-mass limit, they give the two-term formulas for the quantum correction to conductivity \cite{Lu11prl}.
The spin-orbit scattering can be seen as spin-dependent magnetic fields.
It has no effect on the singlet, because singlet state has zero total spin angular momentum of incoming and outgoing scattered electrons.
However, spin-orbit scattering behaves like random magnetic fields to the triplet Cooperon, which carries a net total spin angular momentum of 1.
In this way, the phase coherence length of large-mass Dirac fermions is shortened by spin-orbit scattering, much like by magnetic impurities.

\begin{figure}[htbp]
\centering
\includegraphics[width=0.5\textwidth]{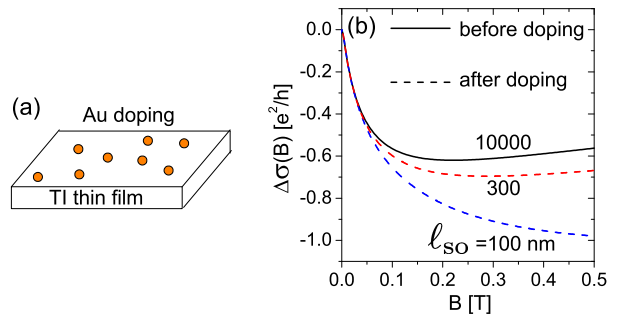}
\caption{(a) A topological insulator thin film doped with impurities (e.g., Au) that bring strong spin-orbit scatterings.
(b) The expected narrowing of the negative magnetoconductivity cusp after Au impurities are doped. In the simulation, one surface band ($\gamma=\pi$) and two bulk subbands ($\gamma=0.15\pi$) are assumed. }
\label{fig:proposal}
\end{figure}

The suppression of WL by spin-orbit scattering provides an experimental signature to detect the WL of bulk states in topological insulators.
It is known that both the bulk and surface states contribute to the transport in as-grown topological insulators \cite{Hor09prb,Qu10sci}.
The experimentally observed negative magnetoconductivity was suggested to be the summation of the WAL from the surface and WL from the two-dimensional bulk subbands \cite{Lu11prb}.
By introducing impurities that can bring spin-orbit scattering, WL from the large-mass bulk states will be suppressed, but WAL from massless surface states will not be affected.
Overall speaking, one will expect a narrowing of negative magnetoconductance cusp as shown in Fig. \ref{fig:proposal}.
The recent experiment on the five quintuple-layer Bi$_2$Te$_3$ thin film with Au impurities may already imply the effect \cite{He11prl}.

\begin{widetext}

\begin{table}[htbp]
\caption{Two-dimensional quantum diffusive transport of conventional and Dirac fermions for impurities of orthogonal (elastic), unitary (magnetic),
and symplectic (spin-orbit) symmetries \cite{Dyson62jmp}. $\tau_m$ and $\tau_ {\mathrm{so}}$ are magnetic and spin-orbit scattering times,
respectively. Elastic scattering ($1/\tau_e\neq 0$) is present in all the cases. $\gamma$ is the Berry phase.}
\label{tab:A}%
\begin{ruledtabular}
\begin{tabular}{ccccc}
&   Orthogonal &  Unitary   & Symplectic \\
&  ($1/\tau_m=1/\tau_{\mathrm{so}}=0$) & ($1/\tau_m\neq 0,1/\tau_{\mathrm{so}}=0$)  &  ($1/\tau_m=0,1/\tau_{\mathrm{so}}\neq0$)\\
  \hline
Conventional electron  &  WL & both suppressed & WL-WAL crossover  \\
Massless Dirac fermion  $\gamma=\pi$ & WAL & suppressed WAL & WAL$^\dag$  \\
Massive Dirac fermion  $\gamma\in (0,\pi)$ & WAL-WL crossover & both suppressed  & WAL-WL crossover$^\dag$\\
Dirac fermion in large-mass limit  $\gamma=0$ & WL & suppressed WL & suppressed WL$^\dag$   \\
\end{tabular}
\end{ruledtabular}
\end{table}

\end{widetext}

To summarize, in Table. \ref{tab:A} we compare the quantum diffusive transport of 2D massive Dirac fermions with the conventional electrons in the presence of impurities of orthogonal (elastic scattering), symplectic (spin-orbit scattering), and unitary (magnetic scattering) symmetries.
The new results in this work are marked by $\dag$ signs.

This work is
supported by the Research Grant Council of Hong Kong under Grant No. HKU 7051/11P.

$^*$ Corresponding author: luhz@hku.hk

$^{\dag}$ Corresponding author: sshen@hku.hk

%\bibliography{refs_wal}

\end{document}